\documentclass[aps,pre,twocolumn,superscriptaddress,showpacs,amssymb,floatfix,a4paper]{revtex4}
\usepackage{graphicx}
\usepackage{graphics}
\usepackage{amsmath}
\usepackage{tabularx}
\usepackage{booktabs}
\usepackage{color}
\usepackage{doi}
\usepackage{textcomp}
\usepackage{comment}
\usepackage[a4paper,margin=1.5cm]{geometry}
\newcolumntype{Y}{>{\centering\arraybackslash}X}

\begin{document}

\title{Critical Short-Time Behavior of Majority-Vote Model on Scale-Free Networks}

\author{D. S. M. Alencar}
\affiliation{Departamento de Física, Universidade Federal do Piauí, 57072-970, Teresina - PI, Brazil}
\author{J. F. S. Neto}
\affiliation{Departamento de Física, Universidade Federal do Piauí, 57072-970, Teresina - PI, Brazil}
\author{T. F. A. Alves}
\affiliation{Departamento de Física, Universidade Federal do Piauí, 57072-970, Teresina - PI, Brazil}
\author{F. W. S. Lima}
\affiliation{Departamento de Física, Universidade Federal do Piauí, 57072-970, Teresina - PI, Brazil}
\author{R. S. Ferreira}
\affiliation{Departamento de Ciências Exatas e Aplicadas, Universidade Federal de Ouro Preto, 35931-008, João Monlevade - MG, Brazil}
\author{G. A. Alves}
\affiliation{Departamento de Física, Universidade Estadual do Piauí, 64002-150, Teresina - PI, Brazil}
\author{A. Macedo-Filho}
\affiliation{Departamento de Física, Universidade Estadual do Piauí, 64002-150, Teresina - PI, Brazil}

\date{Received: date / Revised version: date}

\begin{abstract}

We discuss the short-time behavior of the majority vote dynamics on scale-free networks at the critical threshold. We introduce a heterogeneous mean-field theory on the critical short-time behavior of the majority-vote model on scale-free networks. In addition, we also compare the heterogeneous mean-field predictions with extensive Monte Carlo simulations of the short-time dependencies of the order parameter and the susceptibility. We obtained a closed expression for the dynamical exponent $z$ and the time correlation exponent $\nu_\parallel$. Short-time scaling is compatible with a non-universal critical behavior for $5/2 < \gamma < 7/2$, and for $\gamma \geq 7/2$, we have the mean-field Ising criticality with additional logarithmic corrections for $\gamma=7/2$, in the same way as the stationary scaling.

\end{abstract}



\maketitle

\section{Introduction} \label{sec:introduction}

Recently, we considered the Majority Vote (MV) model\cite{Oliveira-1992, Pereira-2005, Yu-2017, Wu-2009, Crochik-2005, Vilela-2009, Lima-2012, Chen-2015, Vieira-2016, Huang-2017, Krawiecki-2018, Krawiecki-2019, Stanley-2018, Alves-2019} on scale-free networks. Our main result was to extend the stationary heterogeneous mean-field (HMF) theory scaling to include the effects of a degree cutoff needed to preserve the neutral feature of a scale-free network. Unbounded degree fluctuations induce degree correlations, which have, as the main consequence, non-trivial effects on non-equilibrium phase transitions and a deviation from typical HMF scaling\cite{Castellano-2006, Hong-2007, Dorogovtsev-2008}.

Unbounded power-law networks present non-universal scaling in the regime $5/2 < \gamma < 7/2$, where $\gamma$ is the degree distribution exponent\cite{Krawiecki-2019}. HMF scaling for neutral scale-free networks, i.e., after introducing a network cutoff, also leads to a non-universal scaling in the same regime where the critical exponent ratios are functions of the degree exponent $\gamma$\cite{Alencar-2023}. Also, the marginal case $\gamma=7/2$ presents logarithm corrections\cite{Alencar-2023}.

Our primary motivation now is to focus on the short-time dynamics of the MV model. In the critical noise vicinity, we expect the following scaling of the order parameter
\begin{equation}
    M(t) \sim N^{-\beta/\nu_\perp} F\left[ N^{-z}t,N^{1/\nu_\perp}\left(q/q_c-1\right)\right]
\end{equation}
where $q$ is the noise, which plays the role of the control parameter, $q_c$ is the critical threshold, $\nu_\perp$ is the shifting exponent and $z$ is the dynamical exponent, given by
\begin{equation}
    z = \nu_\perp/\nu_\parallel 
\end{equation}
where $\nu_\parallel$ is the correlation time exponent. Short-time dynamics are central in non-equilibrium phase transitions\cite{Henkel-2008}. From the dynamical scaling, we can expect an algebraic decay of the order parameter in the thermodynamic limit in the critical threshold. In the MV model, the noise $q$ acts as a social temperature by giving a concentration of local contrarians.

Short-time dynamics analysis has some advantages\cite{Nascimento-2021}: avoiding critical slowing down, present on the stationary finite-size scaling analysis, and directly accessing the relaxation phenomena of a system, which can present non-trivial features like aging even in non-equilibrium systems\cite{Henkel-2011}. In addition, we can use typical non-equilibrium phenomena, such as the aforementioned algebraic decay, to determine the critical thresholds with good precision. We want to answer whether the ultra-small-world feature of scale-free networks\cite{Cohen-2010, Barabasi-2016} can induce fast relaxation phenomena in consensus formation models.

We applied HMF theory to investigate the critical decaying behavior of the order parameter. We follow an approximate HMF approach to find the short-time scaling. However, we compare our approximate approach with an exact one\cite{Chen-2015, Huang-2017} to verify if the approximate approach would give the correct short-time scaling. Also, we tested our theoretical predictions with short-time Monte Carlo simulations of the MV model on annealed networks with either random regular (RRN) or power-law distributions (PLN) and quenched PLNs generated by the uncorrelated model (UCM)\cite{Catanzaro-2005}. We found that the scale-free network induces fast relaxation of MV dynamics. Moreover, we found a non-universal behavior of the dynamic exponent $z$ in the regime $5/2 < \gamma < 7/2$ by giving theoretical predictions of the $z$ exponent as functions of the degree exponent $\gamma$. 

We organized this paper in the following sections: in Sec.\ \ref{sec:hmfdynamics}, we discuss our HMF results; in Sec.\ \ref{sec:discussion}, we discuss the dynamical scaling, and in Sec.\ \ref{sec:conclusions}, we present our final considerations.

\section{HMF dynamics} \label{sec:hmfdynamics}

We start discussing the dynamical HMF theory of the two-state MV model\cite{Chen-2015, Huang-2017, Alencar-2023}, where we can statistically treat nodes with the same degree as equivalent. In the first moment, we compare different HMF approaches: in one, we obtained an approximate HMF stationary scaling in Ref. \cite{Alencar-2023}, and in another, exact decaying order parameter equations were obtained in Refs. \cite{Chen-2015, Huang-2017}. The two approaches give the same critical threshold on annealed networks, giving the same result when we consider networks with a larger degree. Our objective is to investigate the validity of our approximate approach to determine the dynamical scaling.

Following Ref. \cite{Chen-2015}, we define $p^{+}_k$ as the probability of a node of degree $k$ having a spin on the \textit{up} state. In addition, we can define $P^{+}$ as the probability of a random node having the same \textit{up} state, which obeys the following expression on uncorrelated networks
\begin{equation}
    P^+ = \frac{1}{\langle k \rangle} \sum_{k}p^+_{k}kP(k).
    \label{op}
\end{equation}
We note that $P^+$ should be independent of node degree for a neutral, uncorrelated network and is interpreted as the weighted average fraction of \textit{up} spins.

We can define degree-weighted magnetization as
\begin{equation}
    M^\prime = \frac{1}{\left<k\right>}\sum_{k}k P(k)\sigma_k
    \label{weighted-mag-def}
\end{equation}
where the local stochastic spin variable $\sigma_k$ of a node with degree $k$ can be $\sigma_k = 1$ for \textit{up} spins and $\sigma_k = -1$ for the converse. The relation between $M^\prime$ on Eq. \ref{weighted-mag-def} and $P^+$ on Eq. \ref{op} is readily obtained
\begin{equation}
    M^\prime= 2P^+-1
    \label{weighted-mag-ref}
\end{equation}
In addition, we can write a relation between the standard magnetization $M$ and $p^+_k$
\begin{equation}
    M = 2\sum_{k} P(k)p^+_k - 1.
\end{equation}

We chase exact evolution equations for $M$, $M^\prime$, and $P^+$, which can obtained by the following time evolution equation for the local stochastic spin variable $q_k$\cite{Chen-2015}
\begin{equation}
    \dfrac{d}{dt} p^+_{k} = - p^+_{k}+ \psi_{k},
    \label{time}
\end{equation}
where the first term of the right-hand side represents the spontaneous flip to a \textit{down} state, and the second term on the right-hand side means the flip to the \textit{up} state under the neighborhood influence. The total $\psi_k\left(P^+\right)$ term has two main contributions: either the node agrees with the neighborhood majority with rate $1-q$ or becomes a contrarian with rate $q$
\begin{equation}
    \psi_{k} = (1-q)\varphi_{k} + q\left(1-\varphi_{k}\right).
    \label{psi}
\end{equation}
Now, $\varphi_{k}(P^+)$ is the probability of a $k$-degree neighbor being on the \textit{up} state, given by
\begin{equation}
    \varphi_{k} = \sum_{k^\prime = \left\lceil \frac{k}{2} \right\rceil}^{k}
                  \left(1-\frac{1}{2}\delta_{k^\prime,\frac{k}{2}}\right)
                  \binom{k}{k^\prime}\left(P^+\right)^{k^\prime}\left(1-P^+\right)^{k-k^\prime}.
    \label{varphi}
\end{equation}
Finally, the evolution equation for $P^+$ is obtained by multiplying Eq. \ref{time} by $kP(k)/\langle k\rangle$ and summing in $k$
\begin{equation}
     \dfrac{d}{dt} P^+ = -P^+ + \frac{1}{\langle k \rangle}\sum_{k}kP(k) \psi_{k}.
     \label{master}
\end{equation}

By direct substitution in Eqs. \ref{varphi} and \ref{master}, we note that $P^+ = 1/2$ (paramagnetic state) is a solution irrespective of $q$. Apart from the trivial solution, a linear stability analysis (LSA) around $P^+=1/2$ reveals a critical noise $q_c$ that marks a threshold from the paramagnetic phase to the ferromagnetic one. We make a hypothesis of a weak ferromagnetic solution close to the trivial paramagnetic solution
\begin{equation}
    P^+(t) = \frac{1}{2} + \epsilon(t)
    \label{linearstabilityansatz}
\end{equation}
and seek stable solutions. By inserting Eq. \ref{linearstabilityansatz} on Eq. \ref{master}, we obtain
\begin{equation}
    \dfrac{d}{dt} \epsilon = - \frac{1}{2} - \epsilon + q
                             + \frac{1-2q}{\langle k \rangle}\sum_{k}kP(k) 
                             \varphi_{k}\left(\frac{1}{2}+\epsilon\right),
    \label{stability1}
\end{equation}
and a power series expansion of $\varphi_{k}\left(1/2+\epsilon\right)$ close to $P^+=1/2$, where the zeroth-order term reads
\begin{equation}
    \varphi_{k^\prime}\left(\frac{1}{2}\right) = \sum_{k^\prime= \left\lceil\frac{k}{2}\right\rceil}^{k}
                                        \left(1-\frac{1}{2}\delta_{k^\prime,\left\lceil \frac{k}{2} \right\rceil}\right)
                                        \binom{k}{k^\prime}\left(\frac{1}{2}\right)^{k}.
    \label{stability2}
\end{equation}
and the first-order derivatives read
\begin{equation}
    \dfrac{d\varphi_{k}(x)}{dx} \biggr\rvert_{x = \frac{1}{2}} =
    \left\lbrace
    \begin{aligned}
        & 2^{1-k}k\binom{k}{k/2} \text{ for } k \text{ even;} \\
        & 2^{1-k}k\binom{k-1}{(k-1)/2} \text{ for } k \text{ odd;} \\
    \end{aligned}
    \right.
    \label{stability3}
\end{equation}
yields
\begin{equation}
    \dfrac{d}{dt} \epsilon = \left[-1+\frac{\lambda}{\langle k \rangle}\sum_{k} 
                             k^{2}P(k)2^{1-k}
                             \binom{k-1}{\left\lceil\frac{k-1}{2}\right\rceil}\right]
                             \epsilon,
    \label{stabilityfinal}
\end{equation}
where we defined
\begin{equation}
    \lambda = 1-2q.
\end{equation}

The exponential solution of Eq. \ref{stabilityfinal} gives a growing or decaying average weighted magnetization. The transition point is found when the exponent vanishes
\begin{equation}
    \frac{\lambda_c}{\langle k \rangle}\sum_{k} k^{2}P(k)2^{1-k}\binom{k-1}{\left\lceil\frac{k-1}{2}\right\rceil} = 1,
    \label{critical}
\end{equation}
which is expected to be exact on annealed networks. When a distribution is localized in a specific degree $P(k)=\delta_{k,m}$, as in the case of RRNs, the transition point $q_{c}$ is exact
\begin{equation}
    \lambda^{\text{RRN}}_c = \frac{2^{m-1}}{m} \binom{m-1}{\left\lceil\frac{m-1}{2}\right\rceil}^{-1}.
    \label{critical-noise-rrn}
\end{equation}
Also, in the case of PLNs, we can use the Stirling approximation $k!\approx k^{k+1/2}\sqrt{2\pi}e^{-k}$ for large $k$ to obtain the expression, depending on $P(k)$ moments
 \begin{equation}
    \lambda^{\text{PLN}}_{c} = \sqrt{\frac{\pi}{2}}\frac{\langle k \rangle}{\langle k^{3/2} \rangle}.
    \label{critical-noise-pln}
 \end{equation}
which is the same as our expression for PLNs\cite{Alencar-2023}.

Until here, the approach of Eq. \ref{master} was exact. We can introduce some approximations to investigate HMF scaling and show that they reproduce our HMF scaling\cite{Alencar-2023}. We start by writing Eq. \ref{varphi} as
\begin{equation}
    \varphi_{k} \approx \frac{1}{2} + \frac{1}{2} \text{erf}\left( \sqrt{\frac{k}{2}} M^\prime \right),
    \label{varphi-app}
\end{equation}
which yields, by combining with Eq. \ref{master}
\begin{equation}
    \frac{d}{dt}M^\prime \approx -M^\prime+\frac{\lambda}{\langle k \rangle}\sum_{k}kP(k) \text{erf}{\left( \sqrt{\frac{k}{2}} M^\prime \right)}.
    \label{weighted-mag}
\end{equation}
We can also obtain an approximate time-evolution equation for the standard magnetization in the same way as Eq. \ref{weighted-mag}
 \begin{equation}
    \frac{d}{dt}M \approx -M + \lambda \sum_{k}P(k) \text{erf}{\left( \sqrt{\frac{k}{2}} M^\prime \right)}.
    \label{standard-mag}
 \end{equation}
which also coincides with dynamical evolutions obtained on Ref. \cite{Alencar-2023}, and the HFM stationary scaling is then obtained in the asymptotic $t \to \infty$ limit. 

It is worth remembering that for RRNs, $M$ and $M^\prime$ are the same. The power series expansion on $\text{erf}(x)$ allows us to write the approximate time evolution Eq. \ref{standard-mag} in the case of annealed RRNs as
\begin{equation}
    \frac{d}{dt}M = \varepsilon M - \kappa {M}^{3} + \mathcal{O}({M}^{5}).
    \label{hmf-scaling-final}
\end{equation}
where 
\begin{equation}
    \varepsilon = \frac{\lambda}{\lambda_c} - 1
\end{equation}
and $\kappa$, in the case of RRN distribution, is given by
\begin{equation}
    \kappa^{\text{RRN}}=\frac{\lambda}{3\sqrt{2\pi}}m^{3/2}.
\end{equation}
The solution in the asymptotic limit $M \to 0$ and critical noise $\lambda_c$ ($\varepsilon = 0$) reads
\begin{equation}
    \frac{1}{M(t)^2} - \frac{1}{M(0)^2} \approx 2 \kappa t.
    \label{ode-rr}
\end{equation}
Therefore, we can conclude that $M$ algebraically decays as
\begin{equation}
    \frac{M(t)}{\sqrt{1-M(t)/M(0)}} \propto \left(\kappa t\right)^{-\alpha} \propto \left(\kappa t\right)^{-1/2}
    \label{scaling-time-rrn}
\end{equation}
with $\kappa$ being a constant and the mean-field decaying exponent $\alpha=1/2$. Moreover, for $\lambda<\lambda_c$, we obtain
\begin{equation}
    M(t) \approx \sqrt{\frac{\varepsilon}{\kappa b}} \exp\left( -\left|\varepsilon \right| t \right)
    \label{solution1}
\end{equation}
which is an exponential decay. Also, for $\lambda>\lambda_c$, we obtain
\begin{equation}
    M(t) \approx \sqrt{\frac{\varepsilon}{\kappa b}}\left[ 1 - \frac{b}{2} \exp\left( -\left|\varepsilon \right| t \right) \right]
    \label{solution2}
\end{equation}
where $b$ is an integration constant. Therefore, we have the following relaxation time
\begin{equation}
    \tau = \frac{1}{\left|\varepsilon \right|}.
    \label{relaxationtime}
\end{equation}
From the stationary HMF scaling\cite{Alencar-2023}
\begin{equation}
    \varepsilon \propto N^{-1/\nu_\perp} \propto N^{-1/2}
\end{equation}
which the mean-field exponent $\nu_\perp=2$, we can also write for RRNs
\begin{equation}
    \tau \propto N^{z} \propto N^{1/2}
    \label{relaxationtimescalingrrn}
\end{equation}
which also gives the mean-field exponents $z=1/2$ and $\nu_\parallel = z\nu_\perp = 1$. Finally, from the stationary asymptotic limit of $M(t)$ in Eq. \ref{solution2}, we reobtain the stationary HMF scaling\cite{Alencar-2023}
\begin{equation}
    M \propto N^{-\beta/\nu_\perp} \propto N^{-1/4}
    \label{stationarymagrrn}
\end{equation}
which gives the mean-field exponent $\beta=1/4$. Therefore, for the RRNs, the MV dynamics exhibit mean-field Ising criticality.

For the PL distributions $P(k) \sim k^{-\gamma}$, we should insert the degree cutoff 
\begin{equation}
    k_{c} = N^{1/2},
    \label{cutoff}
\end{equation}
to preserve the neutral feature of the scale-free networks. Uncorrelated degree dependencies would lead to degree correlations and non-trivial effects on phase transitions\cite{Castellano-2006, Hong-2007, Dorogovtsev-2008}. The thermodynamic limit is obtained when $k_{c}$ and $N$ are taken to infinity. On bounded PLN networks with a degree distribution
\begin{equation}
    P(k) = \begin{cases}
              \frac{\gamma-1}{f(\gamma-1)}m^{\gamma-1}k^{-\gamma},  & \quad \text{if } m \leq k \leq k_c; \\
              0,                                                    & \quad \text{if } k < m \text{ and if } k > k_c; \\
           \end{cases}
    \label{cutoffdegreedistribution}
\end{equation}
where
\begin{equation}
    f(x) = 1 - \left( \frac{m}{k_c} \right)^x,
    \label{correction-f}
\end{equation}
we get the same Eq. \ref{hmf-scaling-final} by following Ref. \cite{Alencar-2023}, with $\kappa$ now given by
\begin{equation}
    \kappa^{\text{PLN}} = \frac{g}{6 \lambda a^2}\sqrt{\frac{\pi}{2}}
\end{equation}
where\cite{Alencar-2023}
\begin{equation}
    g = \frac{\left<k^{5/2}\right>}{\left<k\right>} = \frac{\gamma-2}{\gamma-7/2}\frac{f(\gamma-7/2)}{f(\gamma-2)}m^{3/2}
    \label{correctionfactor1}
\end{equation}
and
\begin{equation}
    a = \left< k^{1/2} \right> = \frac{\gamma-1}{\gamma-3/2}\frac{f(\gamma-3/2)}{f(\gamma-1)} m^{1/2}.
    \label{correctionfactor2}
\end{equation}
where $f{x}$ is in Eq. \ref{correction-f}.

The HMF dynamical solutions for $M(t)$ for PLNs are all the same as RRNs, while PLNs now have an extra $g$ dependence, which plays the role of a scaling correction\cite{Alencar-2023}. The scaling decay of the order parameter in the critical threshold for PLNs is
\begin{equation}
    \frac{M(t)}{\sqrt{1-M(t)/M(0)}} \propto \left(\frac{gt}{a^2}\right)^{-1/2}.
    \label{scaling-time-pln}
\end{equation}
The relaxation time is obtained from the stationary scaling of $\left|\varepsilon \right|$, given in Ref. \cite{Alencar-2023}, reading for PLNs
\begin{equation}
    \tau = \frac{1}{\left|\varepsilon \right|} \propto \left( \frac{aN}{g} \right)^{1/2}.
    \label{relaxationtimescalingpln}
\end{equation}
Moreover, from the asymptotic $t\to \infty$ limit of Eq. \ref{solution2}, we obtain the stationary order parameter scaling for PLNs
\begin{equation}
    M \propto \left( \frac{gN}{a^3} \right)^{-1/4}.
    \label{stationarymagpln}
\end{equation}
The same result of Eq. \ref{stationarymagpln} for the stationary scaling can be found by applying the asymptotic limit $M \ll 1$ and make $t=\tau$ in the critical short-time dependence of Eq. \ref{scaling-time-pln}.

In the next section, we will compare the exact approach of Refs. \cite{Chen-2015, Huang-2017} with the approximate HMF approach of this paper, which is necessary to justify posterior results of short-time HMF scaling. Moreover, we compare the HMF predictions with simulation results. We give special attention to short-time simulations to confirm the dynamical exponent $z$ and the time-correlation length exponent $\nu_\parallel$.

\section{Discussion} \label{sec:discussion}

The MV model dynamics are simulated by attempting local spin flips on random nodes with a rate
\begin{equation}
    w(\sigma_{i}) = \frac{1}{2}\left[1-\left(1-2q\right)\sigma_{i}S\left(\sum_{\left<i,j\right>} \sigma_{j} \right)\right]
\end{equation}
where the summation on the right side is done in the first neighbors, and $S(x)$ is the \textit{sign} function, which yields the neighborhood majority opinion
\begin{equation}
S(x) = \begin{cases}
          -1, & \text{if } x < 0; \\
          0,  & \text{if } x = 0; \\
          1,  & \text{if } x > 0. \\
       \end{cases}
\label{signfunction}
\end{equation}

During the dynamics, we collect a time series of the opinion balance $o$, analogous to the magnetization of magnetic equilibrium systems
\begin{equation}
    o = \left\vert \frac{1}{N} \sum_i \sigma_i \right\vert.
\end{equation}
One can calculate the order parameter by averaging $m$ from the opinion balance. In quenched networks, we also do a quenched average on random network realizations. For each random realization, one should evolve dynamics to a stationary state and then collect an ensemble composed of a time series. The order parameter $M$ and the susceptibility $\chi$ are given by the following relations, respectively \cite{Oliveira-1992}
\begin{eqnarray}
M    &=& \left[ \langle o \rangle \right], \nonumber \\
\chi &=& \left[ N (\langle o^{2} \rangle - \langle o \rangle ^{2}) \right]
\label{observables}
\end{eqnarray}
where the symbol $\langle ... \rangle$ represents the average of a time series and the symbol $\left[ ... \right]$ represents the quench average. All collected observables are functions of $\lambda$, and we make averages on stochastic trajectories for short-time simulations. 

\begin{figure*}[!ht]
    \begin{center}
    \includegraphics[scale=0.5]{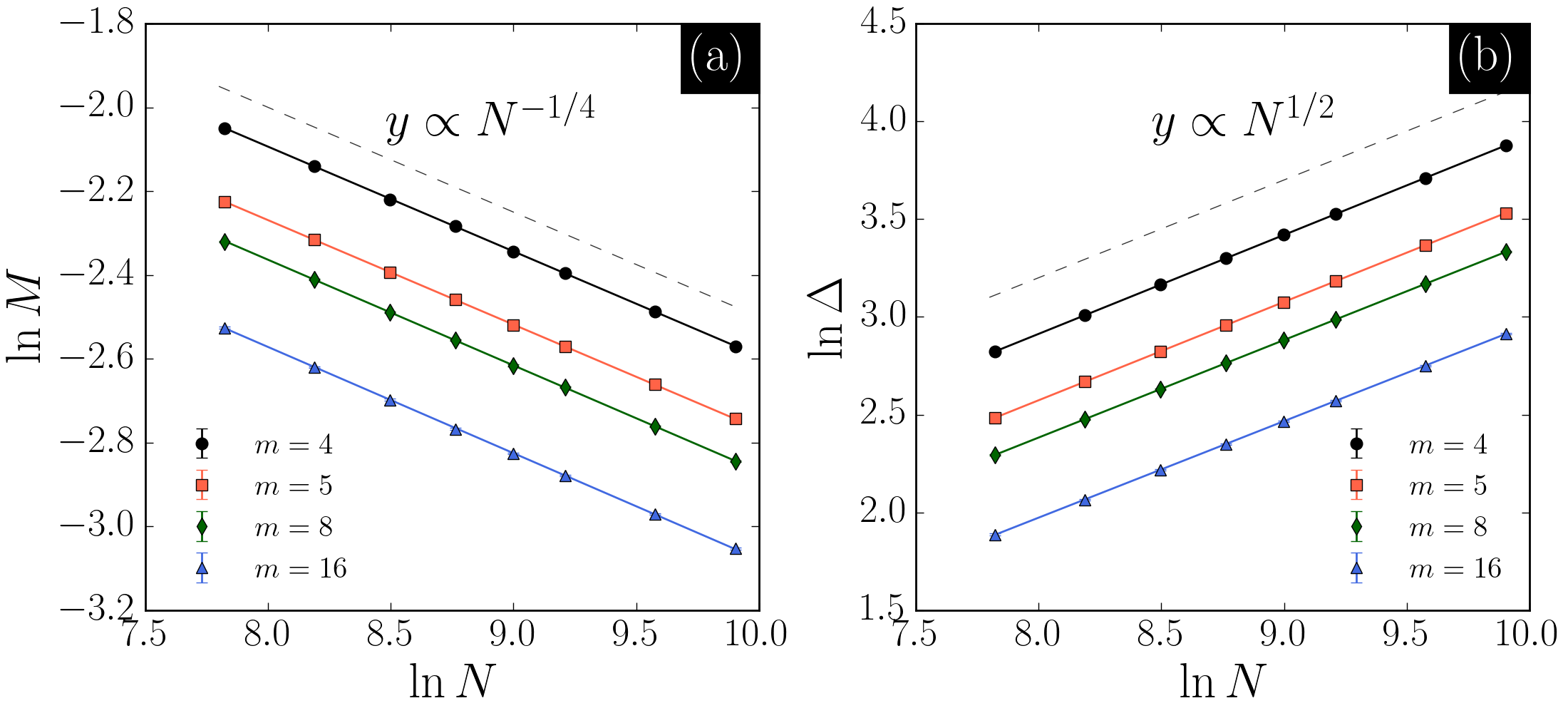}
    \end{center}
    \caption{(Color Online) In panel (a), we show the steady-state decay of the stationary order parameter $M$ as a function of the system size $N$ of annealed random regular networks (RRNs). We simulated $M$ at the critical noise $q_{c}$ for each coordination number $m$, given in Eq. \ref{critical-noise-rrn}. The order parameter algebraically decays at $q_c$ according to Eq. \ref{stationarymagrrn}. In panel (b), we show the stationary susceptibility $\chi$ as a function of the system size $N$ of annealed RRNs. We also simulated the susceptibility data at $q_{c}$. The susceptibility algebraically diverges at $q_{c}$. In both panels, solid lines are regressions, and dashed ones represent mean-field theoretical predictions of the critical exponents for both the order parameter and the susceptibility: $\beta/\nu_\perp = 1/4$ and $\gamma^\prime/\nu_\perp = 1/2$, respectively. All regressions furnish values of the stationary critical exponents that deviate less than $10^{-2}$ from the theoretical values.}
    \label{rrn-regressions}
\end{figure*}

Annealed RRNs provide a way to test the theoretically predicted thresholds of Eq. \ref{critical-noise-rrn} and the HMF critical scaling. In Fig. \ref{rrn-regressions}, we show the stationary behavior of the order parameter $M$ and the susceptibility $\chi$ for annealed RRNs. We waited $10^6$ Monte Carlo steps for the system to reach the stationary state and collected $10^7$ terms of the time series. At each data point, we averaged over $24$ replicas. As expected, the order parameter algebraically decays at $q_c$ according to Eq. \ref{stationarymagrrn} with mean-field exponent $\beta/\nu_\perp = 1/4$. Moreover, the susceptibility $\chi$ algebraically diverges as
\begin{equation}
    \chi \propto N^{-\gamma^\prime/\nu_\perp}
    \label{stationarychirrn}
\end{equation}
at $q_{c}$ with mean-field exponent $\gamma^\prime/\nu_\perp = 1/2$.

\begin{figure*}[!ht]
    \begin{center}
    \includegraphics[scale=0.5]{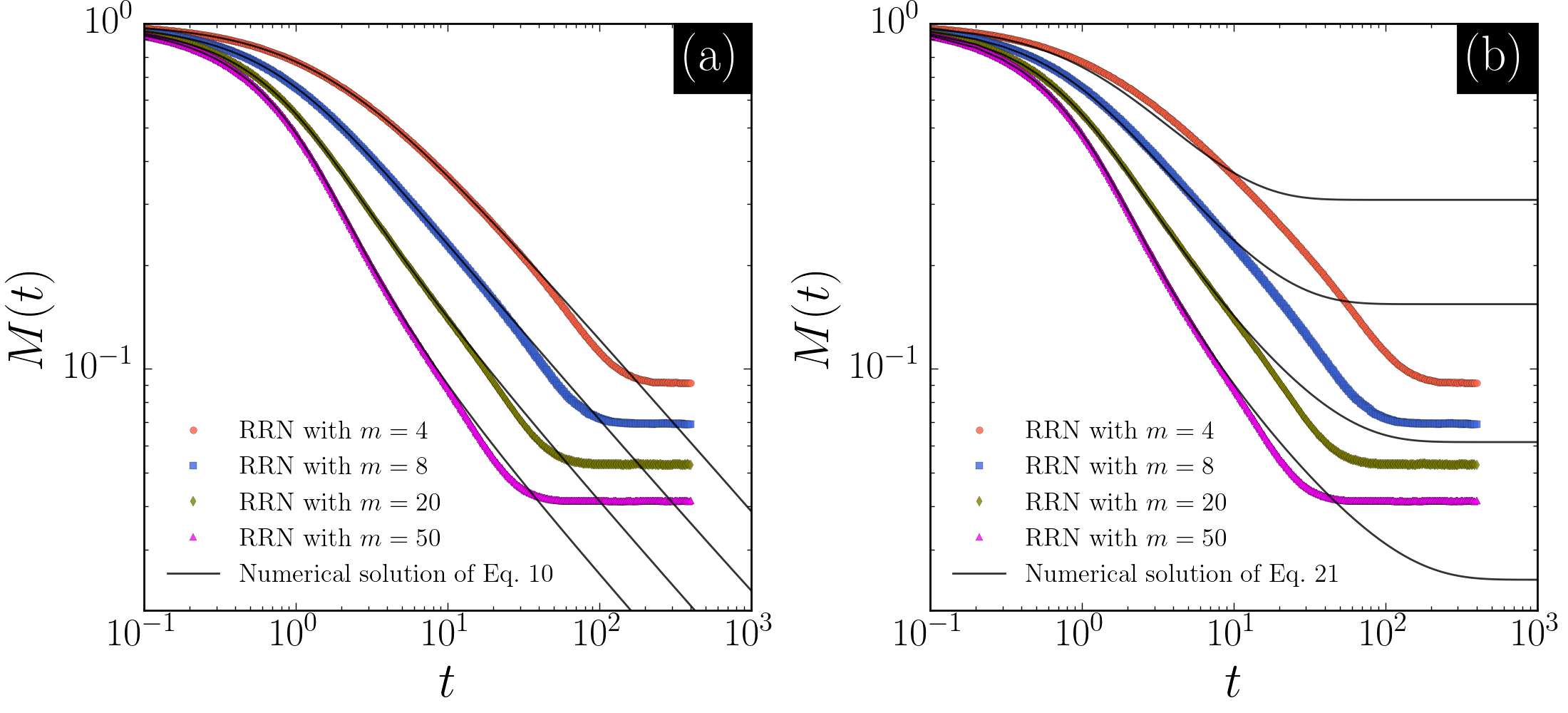}
    \end{center}
    \caption{(Color Online) In all panels, we show short-time simulations of the order parameter $M(t)$ of annealed regular random networks with size $N=10^4$. In all cases, we started from a full magnetized system. We simulated $M$ at the critical noise $q_{c}$ for each coordination number $m$, given in Eq. \ref{critical-noise-rrn}. In panel (a), we compare simulation data with the numerical solution of Eq. \ref{master}, representing the exact approach of Refs. \cite{Chen-2015, Huang-2017}. In panel (b), we do the same as in panel (a), except for the numerical solution of Eq. \ref{weighted-mag}, representing our approximate approach. Both approaches give short-time behavior with reasonable accuracy, which improves when connectivity is increased. We used the approximate approach to determine the dynamic HMF scaling and furnish asymptotic expressions for the critical dynamic exponents.}
    \label{shorttimeapproxvsexact}
\end{figure*}

We also can use RRNs to compare the exact approach of Refs. \cite{Chen-2015, Huang-2017} with the approximate HMF approach of this paper. We show short-time simulations of the order parameter on RRNs and compare them with numerical solutions of the exact Eq. \ref{master} in Fig. \ref{shorttimeapproxvsexact} and the approximate Eq. \ref{weighted-mag}. Each curve results from an average over $10^3$ stochastic trajectories starting from a complete magnetized network ($M(0)=1$) and $160$ realizations of the annealed network. Both approaches successfully predict the correct short-time behavior of the order parameter, which justifies further discussion in finding the correct dynamical HMF scaling and the dynamical exponents, as done in Sec. \ref{sec:hmfdynamics}. We used the 4th-order Runge-Kutta technique to find all the numerical solutions.

\begin{figure*}[!ht]
    \begin{center}
    \includegraphics[scale=0.45]{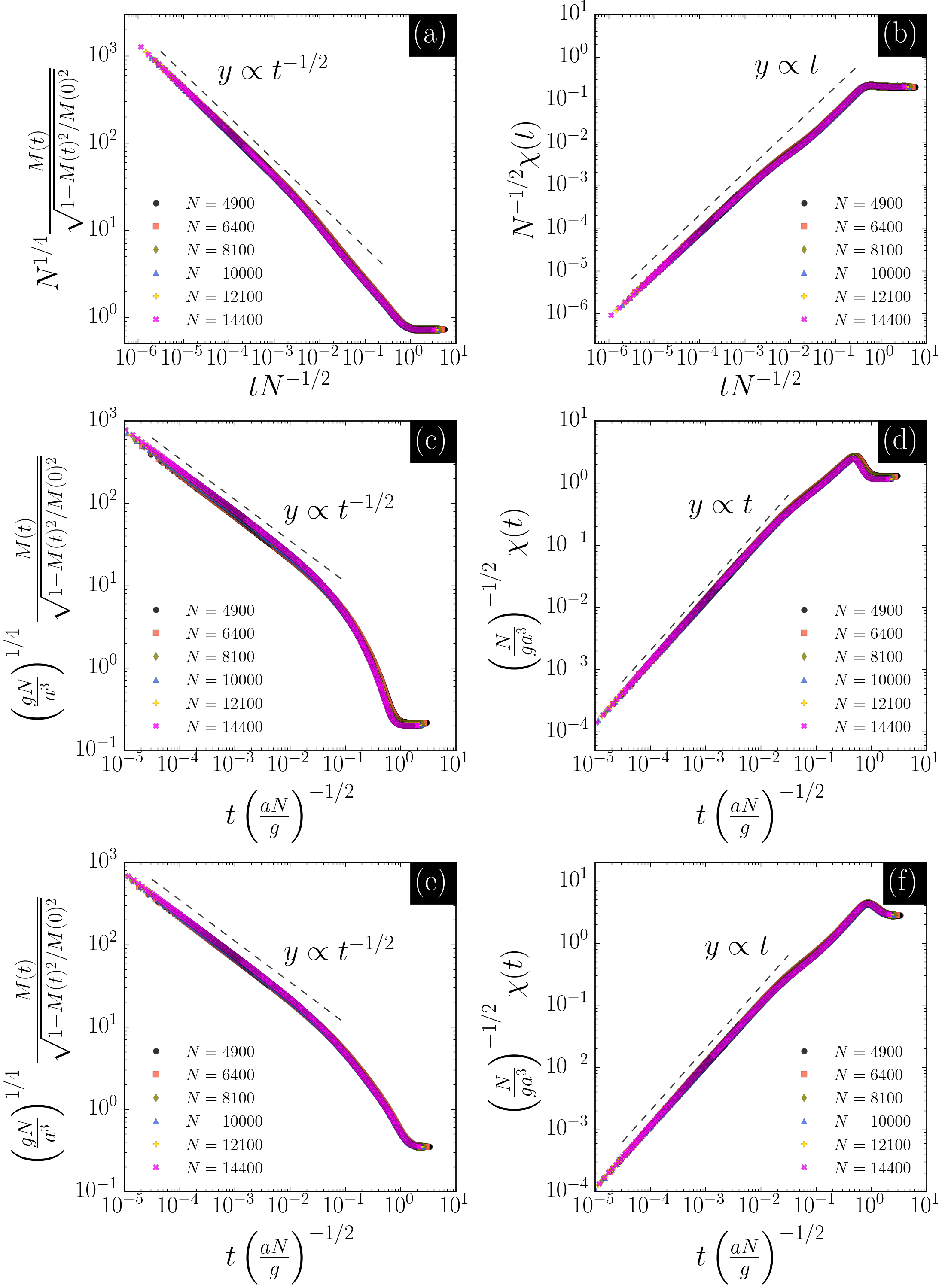}
    \end{center}
    \caption{(Color Online) In panel (a), we show a rescaled plot of the order parameter $M$ \textit{versus} time from short-time simulations of the majority-vote (MV) model on RRNs with some sizes $N$ for the critical noise $q_c \approx 0.271428$. The short-time behavior of $M$ obeys heterogeneous mean-field (HMF) scaling, according to Eq. \ref{scaling-time-rrn}. Also, $M$ obeys the stationary scaling in the final \textit{plateau}, given in Eq. \ref{stationarymagrrn} and the relaxation time scales as Eq. \ref{relaxationtimescalingrrn} as seen from the x-axis rescaling. In panel (b), we show the rescaled susceptibility $\chi$ for the same critical noise as in panel (a), which obeys HMF scaling with the same relaxation time as the order parameter and stationary scaling given by Eq. \ref{stationarychirrn}. In panel (c), we show the same as in panel (a), where we simulated MV dynamics in the critical noise $q_c \approx 0.389221$ on annealed power-law networks (PLNs). The HMF short-time behavior is given in Eq. \ref{scaling-time-pln}, similar to the RRN except with a proportionality constant depending on the correction factor $g$, written in Eq. \ref{correctionfactor2}. The correction factor $g$ scales with the system size, and the scaling dependence leads to faster decaying when decreasing the degree distribution exponent $\gamma$. In addition, data agrees with the stationary scaling on Eq. \ref{stationarymagpln} and the relaxation time scales as Eq. \ref{relaxationtimescalingpln}. In panel (d), we show the same as in panel (b), which obeys HMF short-time scaling with the same relaxation time as the order parameter and stationary scaling given by Eq. \ref{stationarychipln}. In panels (e) and (f), we show the same as panels (a) e (b), respectively; however, for uncorrelated model (UCM)\cite{Catanzaro-2005} PLNs with $\gamma=7/2$ where the critical noise is $q_c \approx 0.3465$\cite{Alencar-2023}. Quenched neutral networks also obey HMF short-time scaling. In all curves, we start from a complete magnetized system, i.e., $M(0)=1$.}
    \label{datacollapses}
\end{figure*}

Regarding the short-time scaling, we now compare the HMF theoretical predictions of Sec. \ref{sec:hmfdynamics} with short-time Monte Carlo simulations. In Fig. \ref{datacollapses}, we show data collapses of the short-time dependence of the order parameter $M(t)$ and the susceptibility $\chi(t)$ on RRNs, PLNs, and quenched PLNs generated by the UCM\cite{Catanzaro-2005}. The UCM networks with a bounded degree distribution $P(k)$ given in Eq. \ref{cutoffdegreedistribution}, can be generated\cite{Catanzaro-2005} by drawing a random degree from the following expression
\begin{equation}
  k_i = \left\lfloor m \left(1 - \frac{k_c^{\gamma-1}-m^{\gamma-1}}{k_c^{\gamma-1}}r\right)^{-1/(\gamma-1)}\right\rfloor.
\end{equation}
where $r$ is a uniformly distributed random number in the $[0,1]$ interval. Now, we assign to each node $i$, in a set of $N$ nodes, $k_i$ stubs. These stubs are $k_i$ incomplete edges from the node $i$ to an unselected neighbor. We also impose the constraint that the sum of all $k_i$ degrees must be even to make an undirected graph. We complete the edges by randomly connecting pairs of stubs from different nodes, forbidding multiple connections. This construction results in a random network with a degree distribution $P(k)$ according to Eq. \ref{cutoffdegreedistribution}. Annealed PLNs are made by just rewiring the stubs in every Monte Carlo pass of the MV dynamics.

In panel (a) of Fig. \ref{datacollapses}, we confirm the short-time HMF scaling for RRNs, summarized in Eq. \ref{scaling-time-rrn}, and also the stationary scaling as expressed in Eq. \ref{relaxationtimescalingrrn}. In addition, in panel (b), we also found that the susceptibility $\chi$ possesses a short-time scale
\begin{equation}
    \chi(t) \propto t
\end{equation}
and the final plateau obeys the same stationary scale of Fig. \ref{rrn-regressions}, given in Eq. \ref{stationarychirrn}, as expected. Each curve of panels (a) and (b) results from an average over $10^3$ stochastic trajectories starting from a full magnetized network and $160$ realizations of the annealed network.

Continuing the discussion regarding the simulations, in panels (c) and (d) of Fig. \ref{datacollapses}, we show results of critical short-time dynamics on annealed PLNs for the order parameter $M(t)$ and the susceptibility $\chi(t)$, respectively. Each curve results from an average over $10^4$ stochastic trajectories starting from a complete magnetized network and $160$ realizations of the annealed network. Data collapses also agree with the HMF short-time behavior, given in Eq. \ref{scaling-time-pln}, where we note the dependence of the relaxation time on the correction factor $g$ written in Eq. \ref{correctionfactor2}. 

The correction factor asymptotically scales with the system size as\cite{Alencar-2023}
\begin{equation}
    g \sim \begin{cases}
            N^{\frac{\gamma-7/2}{\omega}}, & \quad \text{if } 5/2 < \gamma < 7/2; \\
            \ln N,                         & \quad \text{if } \gamma = 7/2; \\   
            \text{constant},               & \quad \text{if } \gamma > 7/2;
    \end{cases}
    \label{correctionfactorscaling}
\end{equation}
where we expect a decreasing relaxation time when decreasing the degree distribution exponent $\gamma$, which leads to faster relaxation. Moreover, the stationary scaling of the order parameter agrees with HMF scaling on Eq. \ref{stationarymagpln}. In panel (d), we can also conclude that the stationary scaling of the susceptibility $\chi$ agrees with
\begin{equation}
    \chi \propto \left( \frac{N}{ga^3} \right)^{1/2},
    \label{stationarychipln}
\end{equation}
which is the same stationary scaling of Ref. \cite{Alencar-2023}. From the stationary scaling being the result of the short-time one with $t=\tau$, we can deduce the short-time scaling as
\begin{equation}
    \chi(t) \propto \frac{t}{a^2}.
\end{equation}
for PLNs. Therefore, the short-time scaling of the susceptibility $\chi$ for PLNs is asymptotically the same as the RRNs.

Finally, in panels (d) and (e) of Fig. \ref{datacollapses}, we show the simulation results of the short-time dynamics on UCM networks. Each curve results from an average over $10^4$ stochastic trajectories starting from a whole magnetized network and $160$ realizations of the annealed network. We aim to see if the quenched uncorrelated PLNs also obey HMF scaling theoretically obtained for annealed ones. Indeed, the short-time and stationary scales of the order parameter $M$ and the susceptibility $\chi$ for quenched UCM networks are compatible with HMF predictions. 

Now, we turn our attention to the dynamic exponents. The relaxation time $\tau$ scales as Eq. \ref{relaxationtimescalingpln}, as seen from the results of panels (c) to (f) of Fig. \ref{datacollapses} where the x-axis is rescaled by $t/\tau$. From the HMF scaling
\begin{equation}
    \tau \propto N^z \propto (gN)^{1/2}
\end{equation}
and the asymptotic $g$ scaling in Eq. \ref{correctionfactorscaling}, we can conclude
\begin{equation}
    z = \begin{cases}
            \frac{1}{2} - \frac{\gamma-7/2}{2\omega}, & \quad \text{if } 5/2 < \gamma < 7/2; \\
            \frac{1}{2},                              & \quad \text{if } \gamma \geq 7/2;
    \end{cases}
    \label{dynamicalexponent}
\end{equation}
with additional logarithmic corrections at $\gamma = 7/2$. The dynamical exponent $z$ equals $1/\nu_\perp$ in Ref. \cite{Alencar-2023}. Therefore, we also obtain
\begin{equation}
    \nu_\parallel = z\nu_\perp = 1.
\end{equation}
The time correlation exponent $\nu_\parallel$ for PLNs is the same as RRNs, supported by HMF theory and simulations. 

\section{Conclusions} \label{sec:conclusions}

We obtained the order parameter's short-time dependence and the MV model's susceptibility on uncorrelated scale-free networks. We also confirmed the stationary scaling of our previous work\cite{Alencar-2023} by the asymptotic limit $t \to \infty$ of the approximate time-dependent evolution equation for the order parameter. We obtained a closed expression for the dynamical exponent $z$ and found the time correlation exponent $\nu_\parallel = 1$.

Critical short-time results also support a non-universal critical behavior for $5/2 < \gamma < 7/2$, where the critical exponents depend on the degree exponent $\gamma$. In the case of $\gamma \geq 7/2$, we obtain asymptotically fixed values of the exponents on the mean-field Ising universality class. However, the majority-vote model presents additional logarithmic corrections for $\gamma=7/2$ on the short-time dynamics of PLNs, in the same way as stationary scaling\cite{Alencar-2023}.

\section*{Acknowledgments} \label{sec:acknowledgements}

We would like to thank CAPES (Coordenação de Aperfeiçoamento de Pessoal de Nível Superior), CNPq (Conselho Nacional de Desenvolvimento Científico e tecnológico),  and FAPEPI (Fundação de Amparo à Pesquisa do Estado do Piauí) for the financial support. We acknowledge the \textit{Dietrich Stauffer Computational Physics Lab.}, Teresina, Brazil, where all computer simulations were performed.

\bibliography{textv1}

\end{document}